# Financial Inclusion and Monetary Policy: A Study on the Relationship between Financial Inclusion and Effectiveness of Monetary Policy in Developing Countries

Gautam Kumar Biswas[1] and Faruque Ahamed[2]

## Abstract

The study analyzed the impact of financial inclusion on the effectiveness of monetary policy in developing countries. By using a panel data set of 10 developing countries during 2004-2020, the study revealed that the financial inclusion measured by the number of ATM per 100,000 adults had a significant negative effect on monetary policy, whereas the other measure of financial inclusion i.e. the number of bank accounts per 100,000 adults had a positive impact on monetary policy, which is not statistically significant. The study also revealed that foreign direct investment (FDI), lending rate and exchange rate had a positive impact on inflation, but only the effect of lending rate is statistically significant. Therefore, the governments of these countries should make necessary drives to increase the level of financial inclusion as it stabilizes the price level by reducing the inflation in the economy.

## Introduction

To achieve sustainable economic growth is the ultimate target of every nation around the world. Researchers have identified several factors such as export earnings, foreign aid, foreign direct investment, remittances etc. contributing for such growth (Balassa, 1978; Islam & Biswas, 2023; Ranjan & Agrawal, 2011). Most recently researchers, academics, and politicians around the world, especially in developing nations, are finding that financial inclusion is an increasingly fascinating topic. The term "financial inclusion" refers to people having simple access to valuable and reasonably priced financial products and services to suit a variety of requirements, including transactions, payments, and savings. It has become a key policy goal for several developing countries to promote inclusive economic expansion as well as reduce poverty. Monetary policy, on the other hand, is a crucial tool for managing a country's economy. However, there is still a lack of knowledge regarding the connection between financial inclusion and the efficiency of monetary policy, especially in developing nations where achieving both is still difficult. The study aims to measure the effect of financial inclusion on the effectiveness of monetary policy in ten developing countries. The study is organized into the following sections:

[1] Ph.D. Candidate, Department of Economics, Southern Illinois University-Carbondale
[2] MSc in Economics, Northern Illinois University

Section 1 introduces the topic, Section 2 states the research question, Section 3 presents the literature review, and Section 4 describes the data and the model. In Section 5, the empirical results are explained, Section 6 provides a summary of the research along with suggestions, and Section 7 displays the reference section of the study.

## Research Objective

This study aims to examine the relationship between financial inclusion and the effectiveness of monetary policy in developing countries. In particular, the study aims to answer the following research questions:

- Does financial inclusion affect the effectiveness of monetary policy in developing countries?

## Literature Review

For developed nations, the subject of financial inclusion is well-known, but for developing nations, it is a new area of study. In developed countries, there has been extensive research on financial inclusion and its effect on monetary policy, however, there is a severe lack of literature on this topic for underdeveloped countries. This paper attempts to investigate how financial inclusion affects the efficacy of monetary policy in emerging nations in this study.

From 2004 to 2013, the South Asian Association for Regional Cooperation (SAARC) countries were the focus of Lenka & Bairwa's (2016) investigation of the effect of financial inclusion on monetary policy. The authors used a Financial Inclusion Index in their study, which has been constructed through principal component analysis. They discovered a bad correlation between inflation, interest rates, currency rates, and financial inclusion in SAARC nations.

Mehrotra & Yetman (2014) explored how financial inclusion, or having access to regulated financial services, affects optimal monetary policy. The theoretical framework of their study was based on the idea that only financially stable households can borrow money and save money to stabilize their spending during periods of income fluctuation. The study found a correlation between the proportion of households with access to financial services and the ratio of output volatility to inflation volatility suggested by optimal monetary policy. Anjom, & Faruq (2023) analyzed the key factors of financial stability of listed Shariah-based Islamic Banks of Bangladesh using bank-specific, bank industry-specific, and macroeconomic factors. The study finds that the financial stability of the banking system can be affected by macroeconomic variables. Medium- and long-term financial stability in the banking system has positive relationship with the financial inclusion. Faruq (2023) further concluded that foreign direct investments also have positive link with the monetary policy affecting the financial inclusion in the long run.

Thi Truc Nguyen (2018) explored the relationship between financial inclusion and monetary policy in Vietnam using PCA method based on Comprehensive Financial Inclusion Index. The study revealed a negative financial inclusion's effect on monetary policy, indicating that efficient financial intermediation contributes to a stable and sustainable economy. The study also came to the conclusion that financial inclusion might help decision-makers forecast inflationary trends more accurately. Saraswati et al., (2020) investigated the link between Indonesian monetary policy effectiveness, fintech, and financial inclusion. According to the authors, financial inclusion has a short- and long-term impact on inflation rates, but these effects are only temporary. Fintech only has a short-term impact on inflation rates, according to the report. However, through substitution and cost of capital effects, fintech shocks have a long-lasting impact on the volatility of the inflation rate. Arshad et al., (2021) analyzed using structural vector auto-regressive methods the relationship between monetary policy efficacy and financial inclusion in developed and undeveloped nations. The study indicated that while efficient monetary policy improves financial inclusion in a nation, the two do not necessarily have a simultaneous impact on one another. The study also finds a one-way causal relationship between financial inclusion in developing nations and the effectiveness of monetary policy. (Barik & Pradhan, 2021) examined the association between financial inclusion and financial stability among the BRICS countries from 2005 to 2015. Financial inclusion, according to the authors, undermines financial stability. They also concluded that financial inclusion has a detrimental impact on financial stability due to variables like rapid loan expansion, deterioration of credit standards, and insufficient banking sector regulation.

## Data and Model Specification

The aim of this study is to examine the role of financial inclusion on monetary policy effectiveness in developing countries. The data will be derived from the International Monetary Fund's Financial Access Survey (FAS) and the World Bank's World Development Index (WDI) database. Inflation will be included as proxy variable for monetary policy as it reflects the overall macroeconomic condition of a country (Barro, 1995; Biswas, 2023). Two measures of financial inclusion are considered in this study. The availability dimensions are calculated using the natural logarithm of the number of bank branches for 100,000 adults and the natural logarithm of the number of ATMs for 100,000 individuals. These measures are important as transaction points are essential in a comprehensive financial system and these should be easily accessible to users (Sarma, 2016). Monetary policy generally aims to maintain price stability, ensure employment generation, and achieve economic growth. The foremost and important role of monetary policy is to control inflation and ensure stability in general price levels. Therefore, in my study I considered inflation a proxy for assessing the success of monetary policy and I used lending rate (LR), exchange rate (EX) and Foreign Direct Investment (FDI) as other control variables. Thus, the regression to be estimated is the following:

$$INF_{it} = \beta + \alpha_1 FI_{it} + \alpha_2 FDI_{it} + \alpha_3 EX_{it} + \alpha_2 LR_{it} + \eta_t + \varepsilon_{it} \qquad (1)$$

where, $INF_{it}$ is inflation, $FI_{it}$ denotes the different financial inclusion measurements, $FDI_{it}$ is foreign direct investment, $EX_{it}$ is the exchange rate, $LR_{it}$ is lending rate, $\eta_t$ is an unobserved country-specific fixed effect and $\varepsilon_{it}$ is the error term.

In my study, several variants of equation (1) will be estimated to explore the effects of financial inclusion on monetary policy variables.

I estimated the pooled model estimation, fixed effect estimation and random effect estimation of equation (1) to explore the effect of financial inclusion on monetary policy effectiveness. I have used inflation as the monetary policy variable.

## Empirical Results

The descriptive statistics of the research variables are presented in Table 1. This shows that there is a large financial gap between the 10 developing countries around the world, shown by the significant difference between the minimum and maximum value of the logarithm of bank branches per 100,000 adult population. The data variation for inflation is also high as shown as the maximum value is 27.28, while the minimum value is -6.81 (see Table 1).

## Table 1: Summary Statistics

| Variable | Mean | Median | Min | Max |
|---|---|---|---|---|
| Log bank branch | 14.2799 | 10.4081 | 0.3666 | 59.1080 |
| Log of ATM | 1.0303 | 1.0201 | -1.8069 | 2.5677 |
| Log of FDI | 9.109 | 9.240 | 6.423 | 11.010 |
| log of lending rate | 1.1714 | 1.1121 | 0.8608 | 1.7434 |
| Log of exchange rate | 1.8086 | 1.7774 | 0.2235 | 4.1638 |
| Inflation | 6.811 | 6.164 | -6.811 | 27.283 |

## Table 2: Pooled Model Estimation

| Variables | Dependent Variable: Inflation | |
|---|---|---|
| | Pooled Model (1) | Pooled Model (2) |
| Intercept | -3.47 | 0.76 |
| | (3.50) | (3.51) |
| Log of ATM | -1.35*** | |
| | (0.49) | |
| Log of Bank Branch | | -2.63** |
| | | (1.12) |
| Log of FDI | 0.03 | -0.07 |
| | (0.35) | (0.34) |
| Log of Lending rate | 8.10*** | 6.52*** |
| | (1.60) | (1.78) |
| Log of exchange rate | 1.10*** | 0.97** |
| | (0.38) | (0.41) |
| Observations | 160 | 161 |
| R-squared | 0.22 | 0.21 |

Standard errors in parentheses, ∗ significant at 10%; ∗∗ significant at 5%; ∗∗∗ significant at 1%.

The pooled model estimation of equation (1) shows that the effects of financial inclusion (measured by average ATM for 100K adults and Average bank branches for 100K adults) on inflation are negative, which are statistically significant. The results of pooled model (1) states that 1 % increases in the number of ATM per 100,000 adults reduces inflation by 0.0135 (1.35/100), while 1 % increases in the number of bank branches per 100,000 adults reduces inflation by 0.0263. But the pooled regression model might be biased and inconsistent due to serial correlation among errors. Hence, in the following section I estimated the fixed effect and random effect model.

The results of fixed effect estimation and random effect estimation of equation (1) are presented in Table 3.

**Table 3: Fixed Effect and Random Effect Estimation**

| Variables | Dependent Variable : Inflation | | | |
|---|---|---|---|---|
| | Fixed Effect (1) | Fixed Effect (2) | Random Effect (1) | Random Effect (2) |
| Intercept | | | -13.83*** (1.83) | -12.38*** (1.82) |
| Log of ATM | -0.43 (0.84) | | -0.86*** (0.19) | |
| Log of Bank Branch | | 2.09 (2.69) | | 0.02 (0.52) |
| Log of FDI | 1.22 (1.01) | 1.02 (0.98) | 0.33 (0.17) | 0.03 (0.18) |
| Log of Lending rate | 22.04*** (4.72) | 23.44*** (4.77) | 13.67*** (0.84) | 13.80*** (0.87) |
| Log of exchange rate | 0.51 (1.23) | 0.52 (1.23) | 1.41 (0.21) | 1.56*** (0.22) |
| Observations | 160 | 161 | 160 | 161 |
| R-squared | 0.15 | 0.16 | 0.19 | 0.17 |

Standard errors in parentheses, ∗ significant at 10%; ∗∗ significant at 5%; ∗∗∗ significant at 1%.

Table 3 shows that the effect of availability of ATM for 100K population on inflation is negative for both fixed effect and random effect model. The estimated coefficient from random effect model is large, which is statistically significant. While the effect of number of availabilities of Bank Branches for 100K population on inflation in both fixed effect model and random effect model are positive, but the results are not statistically significant. The effects of other control variables - FDI, lending rate and exchange rate all have positive effects on inflation in both fixed effect and random effect estimations, but among these the effect of lending rate is statistically significant.

Finally, I conducted the "Hausman Test" to check which model is appropriate for estimation. The hypothesis of Hausman test is as follows:

| | |
|---|---|
| $H_0$ | The Random Effect model is appropriate |
| $H_1$ | The Fixed Effect model is appropriate |

In both cases, I failed to reject the null hypothesis, which implied that the random effect model is appropriate for estimation.

## Conclusion

This study used annual data over the period 2004-2020 for estimating the effect of financial inclusion on monetary policy effectiveness in 10 developing countries. In my study, I used inflation as the proxy for monetary policy effectiveness and two measures of financial inclusion namely – the number of ATM per 100,000 adults and the number of bank branches per 100,000 adults are used as the proxies of financial inclusion. The study has found that the number of ATM per 100,000 adults has negative and significant effect on inflation, while the other measures of financial inclusion i.e., the number of bank branches per 100,000 adults has positive effect on inflation, which is not statistically significant. It indicates that if the financial inclusion increases, then it reduces inflation, which causes price level to be stable. The study also revealed that other control variables, namely FDI, lending rate and exchange rate have positive impact on inflation. Finally, the Hausman test confirms that the random effect model is the appropriate model for estimation purpose.

Based on empirical findings, the governments of these developing countries should make necessary drives to increase the level of financial inclusion as it reduces the inflation, which improves the effectiveness of the monetary policy as it stabilizes the price level in the economy.

The study was limited to data from specific developing countries and in my study, I only used two measures of financial inclusion. In future study, a large number of countries can be included as well as a comprehensive financial index using several measures of different dimension of financial inclusion can be developed to examine the effect of financial inclusion on the effectiveness of monetary policy.

## Statements:

The authors declare no competing interests. The author also declares that there exists no conflict of interest. The author didn't receive any grant or funding for this report. All the data are collected from primary sources with the approval of the relevant Authority.